\pdfoutput=1

\documentclass[11pt]{article}

\usepackage[final]{acl}

\usepackage{times}
\usepackage{latexsym}

\usepackage[T1]{fontenc}

\usepackage[utf8]{inputenc}

\usepackage{microtype}

\usepackage{inconsolata}

\usepackage{graphicx}

\usepackage{tabularray}
\usepackage{longtable}
\usepackage{xcolor}
\usepackage{colortbl}
\usepackage{multicol}
\usepackage{booktabs}
\usepackage{makecell}
\usepackage{amsmath, amssymb}
\usepackage{multirow}
\usepackage{mathtools}
\usepackage[inline]{enumitem}
\usepackage{subcaption}
\usepackage{float}
\usepackage{graphicx}
\usepackage{caption} 
\usepackage{upgreek}
\usepackage{seqsplit}
\usepackage{color,soul}
\usepackage{arydshln}
\usepackage{setspace}

\usepackage[most]{tcolorbox}
\newtcolorbox{mybox}[2][]{
    colback=white,
    colframe=gray!45,
    fonttitle=\bfseries,
    coltitle=black,
    sharp corners,
    title=#2,
    #1
}

\usepackage{amsmath,amsfonts,bm}

\def\eqref#1{equation~\ref{#1}}

\def\1{\bm{1}}

\def\vc{{\bm{c}}}

\DeclareMathAlphabet{\mathsfit}{\encodingdefault}{\sfdefault}{m}{sl}
\SetMathAlphabet{\mathsfit}{bold}{\encodingdefault}{\sfdefault}{bx}{n}

\usepackage{listings}
\usepackage{xcolor}

\definecolor{codegreen}{rgb}{0,0.6,0}
\definecolor{codegray}{rgb}{0.5,0.5,0.5}
\definecolor{codepurple}{rgb}{0.58,0,0.82}
\definecolor{backcolour}{rgb}{0.95,0.95,0.92}

\lstdefinestyle{mystyle}{
    backgroundcolor=\color{backcolour},   
    commentstyle=\color{codegreen},
    keywordstyle=\color{magenta},
    numberstyle=\tiny\color{codegray},
    stringstyle=\color{codepurple},
    basicstyle=\ttfamily\footnotesize,
    breakatwhitespace=false,         
    breaklines=true,                 
    captionpos=b,                    
    keepspaces=true,                 
    numbers=left,                    
    numbersep=5pt,                  
    showspaces=false,                
    showstringspaces=false,
    showtabs=false,                  
    tabsize=2
}

\usepackage{caption}
\usepackage{subcaption}
\title{Rethinking Code Refinement: Learning to Judge Code Efficiency}

\author{
    Minju Seo$^{1}$ \;\;
    Jinheon Baek$^{1}$ \;\;
    Sung Ju Hwang$^{1,2}$ \\
    KAIST$^{1}$, \;\; DeepAuto$^{2}$ \\
    \texttt{\{minjuseo, jinheon.baek, sjhwang82\}@kaist.ac.kr}
}

\begin{document}
\maketitle

\begin{abstract}

Large Language Models (LLMs) have demonstrated impressive capabilities in understanding and generating codes. Due to these capabilities, many recent methods are proposed to automatically refine the codes with LLMs. However, we should rethink that the refined codes (from LLMs and even humans) are not always more efficient than their original versions. On the other hand, running two different versions of codes and comparing them every time is not ideal and time-consuming. Therefore, in this work, we propose a novel method based on the code language model that is trained to judge the efficiency between two different codes (generated across humans and machines) by either classifying the superior one or predicting the relative improvement. We validate our method on multiple programming languages with multiple refinement steps, demonstrating that the proposed method can effectively distinguish between more and less efficient versions of code. 

\end{abstract}
\section{Introduction}

Large Language Models (LLMs) have shown significant success across a wide range of tasks, extending from natural language understanding to programming-related activities~\cite{GPT3,Codex,GPT4,Llama2,CodeLlama,phi3}. Specifically, thanks to their capabilities in understanding and generating codes, LLMs are able to allow developers to save time, reduce errors, and boost their productivity~\cite{JSCodeGen}. For example, several recent studies have utilized LLMs to optimize and refine the existing code bases~\cite{pie-perf, Supersonic}. Also, more recent work has been proposed to iteratively refine the generated codes from LLMs by judging them with LLMs~\cite{STOP}. Another noteworthy approach involves using LLMs to check the functional correctness of the generated codes from LLMs~\cite{CodeScore}.

\begin{figure*}[t!]
    \centering
    \includegraphics[width=0.975\linewidth]{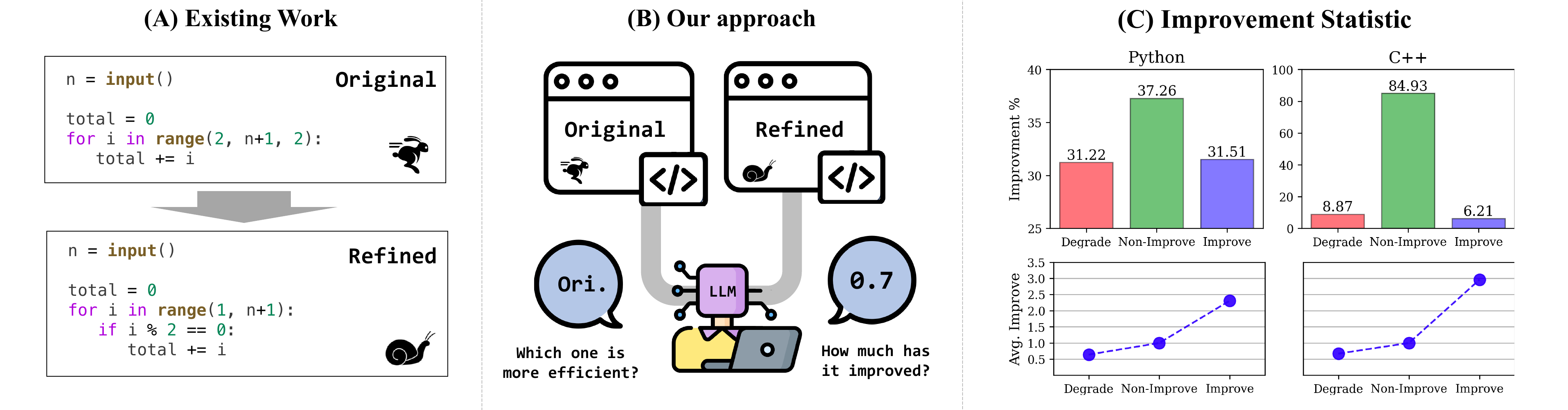}
    \vspace{-0.15in}
    \caption{(A) Existing code refinement approaches sometimes generate the code which has inferior efficiency to the original code. (B) Our proposed approach identifies the efficient code among two different versions of codes (before and after modifications), and further predicts its relative improvement. (C) We categorize the refined code according to its efficiency gain (\%) compared to the original into three classes: Degradation (less than 0.9), Non-Improvement (0.9 to 1.1), and Improvement (greater than 1.1).}
    \label{fig:main_figure}
    \vspace{-0.15in}
\end{figure*}

However, despite huge advancements made in the field of LLMs for code generation, the aforementioned studies assume that the codes generated and refined from LLMs are more efficient than their originals. However, as shown in Figure~\ref{fig:main_figure}, our observations contradict this assumption, showing that LLM-generated and -refined codes do not always perform better. To handle this issue, while one may calculate the efficiencies of codes both before after modifications by actually executing them, this process introduces unnecessary inefficiencies, and may be very costly and time-consuming.

In this work, to overcome those challenges, we introduce a new task of judging the efficiency of the refined code over its original version. In addition, not only LLMs but also human coders may degrade the efficiency of codes during refinement; thus, our task of judging the efficiency between two codes involves all the possible pairs of code modification sources, including human-human, human-machine, and machine-machine. Then, to address this new task, we propose a new model (based on the code LM) that is trained to compare efficiencies between two codes. Specifically, given a code pair (one is original and the other is refined from it), the code LM is trained to classify which one is more efficient or predict how much the refined code is efficient. 

We experimentally validate the effectiveness of our efficiency judgement model on multiple code refinement scenarios with multiple programming languages. The results show that our judgement model can identify the more efficient code among two different versions, substantially outperforming baselines. Our contributions are as follows:

\vspace{-0.075in}
\begin{itemize}[itemsep=0.5mm, parsep=0.5pt, leftmargin=*]
    \item We point out that the refined codes from LLMs or humans do not always improve their efficiency.
    \item We introduce a novel approach that judges the efficiency of two different versions of codes. 
    \item We validate our model on multiple code refinement scenarios, demonstrating its effectiveness.
\end{itemize}

\section{Related Work}

\vspace{-0.05in}
\paragraph{Large Language Models for Code}
Large Language Models (LLMs), trained on extensive corpora with multi-billion parameters, exhibit remarkable performance across a broad spectrum of tasks involving both text and code~\cite{GPT3, AlphaCode, CodeLlama, starcoder, GPT4, phi3, deepseekcoder}. These models, particularly those trained on code-specific datasets, have opened up a new era in software development by not only assisting with basic programming tasks but also enabling more complex activities such as code generation~\cite{CodeT, CodeGen}, translation~\cite{tranx, UnsupTrans, hr2lr}, and refinement~\cite{WaveCoder, RepairT5}. As such, these models are increasingly integrated into development environments, optimizing workflows, and reducing the time for development \cite{JSCodeGen,SelfColabChatGPT}.

\paragraph{LLMs for Code Refinement}
Beyond basic code generation, LLMs are widely used to refine the existing code bases. One of the early work in code refinement aims to detect and fix bugs in the codes by pre-training a transformer-based model on English and source code, and then fine-tuning it on commits (relevant to the part fixing bugs and improving performance)~\cite{DeepPERF}. In a similar vein, another work proposes to refine the code with a sequence-to-sequence model that is trained to transform the original code to its optimized version of the code~\cite{Supersonic}. Additionally, recent work showcases that LLMs are able to recursively improve their own generated codes, progressively enhancing their outputs~\cite{STOP}. Further, \citet{pie-perf} demonstrate that LLMs with sophisticated prompting strategies (to adapt LLM for code optimization) can surpass human-level performance in code optimization tasks. However, despite these substantial achievements, prior studies have primarily focused on code enhancement with limited attention to the actual efficiency of the refined code. Meanwhile, we focus on a different angle, proposing to judge the efficiency of the refined codes in advance (without executing them) based on our observation that not all the refined codes have better efficiency.

\paragraph{LLM-Powered Code Evaluation}
The objective of our work which aims to evaluate the efficiency between two codes based on LLMs has a similarity to work on LLMs for code evaluation. Early work on it uses either a term-matching-based approach (similar to BLEU) or an embedding-based approach (whose representations are obtained from language models), to compare two codes~\cite{CodeBLEU, CodeBERTScore}. However, as collecting ground-truth answers for every evaluation is difficult, recent work has shifted towards using LLMs to judge the quality of the generated code, such as its utility or functional correctness, without the need for comparisons to the reference code~\cite{CodeScore, ICEScore}. Yet, unlike these approaches evaluating the single instance of the code other than the efficiency, we aim to compare efficiency in the setting where the code pair is given.

\section{Method}

We first provide a general description of code refinement, and then introduce our approach.

\subsection{Code Refinement}

Let us assume that the existing code base is defined as $\vc$, which consists of a sequence of tokens as follows: $\vc = \{ c_1, ..., c_n \}$. Then, the objective of code refinement (in this work) is to transform the existing code base $\vc$ into its improved version $\vc' = \{ c'_1, ..., c'_m \}$, where the execution time for $\vc'$ should be faster than $\vc$, formalized as follows: $\texttt{Exec}(\vc) > \texttt{Exec}(\vc')$. Here, $\texttt{Exec}$ is the code execution function that returns its runtime.

It is worth noting that, in this work, we consider three different scenarios of code refinement. The first scenario involves a human-human interaction, where one developer revises the code originally authored by another. The second scenario, termed the human-machine scenario, consists of collaborative efforts between humans and machines. This practice has become increasingly prevalent in real-world software development environments, thanks to CodeLLMs~\cite{Codex, CodeLlama}. Lastly, the machine-machine scenario involves autonomous code refinement by machines, a process that has shown promise in various studies~\cite{STOP, pie-perf}. 

Note that, despite the huge advancements made in the field of code refinement, we find that modified codes from machines can occasionally reduce the efficiency of the original codes. Similarly, human developers may diminish the efficiency of the codes during refinement. On the other hand, executing the pair of original and modified codes at every refinement step is inefficient and time-consuming.

\begin{table}[t!]
\caption{Main results on the task of judging the code efficiency, where easy denotes the dataset containing only the code pairs whose efficiency difference is more than 10\%.}
\vspace{-0.1in}
\label{tab:main_classification}
\small
\centering
\resizebox{0.475\textwidth}{!}{
\renewcommand{\arraystretch}{1.0}
\begin{tabular}{lcccccc}
\toprule

 & \multicolumn{3}{c}{\bf All} & \multicolumn{3}{c}{\bf Easy }  \\
\cmidrule(l{2pt}r{2pt}){2-4} \cmidrule(l{2pt}r{2pt}){5-7}
 & Python & C++ & All & Python & C++ & All \\

\midrule
\midrule

Zero-shot & 50.87 & 45.30 & 46.55 & 47.04 & 51.21 & 49.70  \\
Few-shot & 51.21 & 51.17 & 51.18 & 49.16 & 48.22 & 48.56  \\

\noalign{\vskip 0.25ex}\cdashline{1-7}\noalign{\vskip 0.75ex}

Zero-shot CoT & 50.35 & 51.69 & 51.39 & 48.94 & 52.10 &  50.95 \\
Few-shot CoT & 50.52 & 50.87 & 50.79 & 49.50 & 48.09 & 48.60  \\

\noalign{\vskip 0.25ex}\cdashline{1-7}\noalign{\vskip 0.75ex}

GPT-3.5 & 54.50 & 52.17	 & 52.69 & 54.97 & 55.59 & 55.37 \\
GPT-4o & 63.67 & 56.03 & 57.75 & 66.03 & 60.48 & 62.49 \\

\noalign{\vskip 0.25ex}\cdashline{1-7}\noalign{\vskip 0.75ex}

\textbf{Ours} & \textbf{72.49} & \textbf{62.08} & \textbf{64.42} & \textbf{77.65} & \textbf{70.14} & \textbf{72.86}
 \\

\bottomrule

\end{tabular}
}
\vspace{-0.125in}
\end{table}

\subsection{Judging Code Efficiency}

To overcome the aforementioned limitation, we aim to predict the efficiency of the modified code over its original code, without actually executing them. This can be formulated by either the classification problem (where we classify the superior code) or the regression problem (where we predict the relative improvement of the modified code over the original code), given a pair of original and modified codes. Also, note that we operationalize classification and regression problems with CodeLLMs, due to their capabilities in understanding codes.

Specifically, given the code pair (e.g., $\vc$ and $\vc'$), we concatenate it and provide it with the CodeLLM, formalized as follows: $o = \texttt{CodeLLM}([\vc, \vc'])$ where $[\cdot]$ is the concatenation operation, and $o$ is the prediction output. Then, for the classification problem, we formulate it as the next token prediction task:
$
    L_C=-\log p(o' | [\vc, \vc'], o' \in \{A,B\}),
$
where $o'$ is ground truth that belongs to one of $A$ or $B$, in which $A$ represents the improvement over the original code and vice versa for $B$. Similarly, for the regression problem, we train the model to predict the relative improvement of the refined code over its original code by minimizing this prediction value with the actual relative improvement.

\section{Experiment}
\vspace{-0.025in}

We now describe experimental setups and results. We provide our code at \href{https://github.com/going-doer/judge_code_efficiency}{https://github.com/going-doer/judge\_code\_efficiency}, for reproducibility.

\subsection{Experimental Setups}

\begin{figure}[t!]
    \centering
    \includegraphics[width=0.95\linewidth]{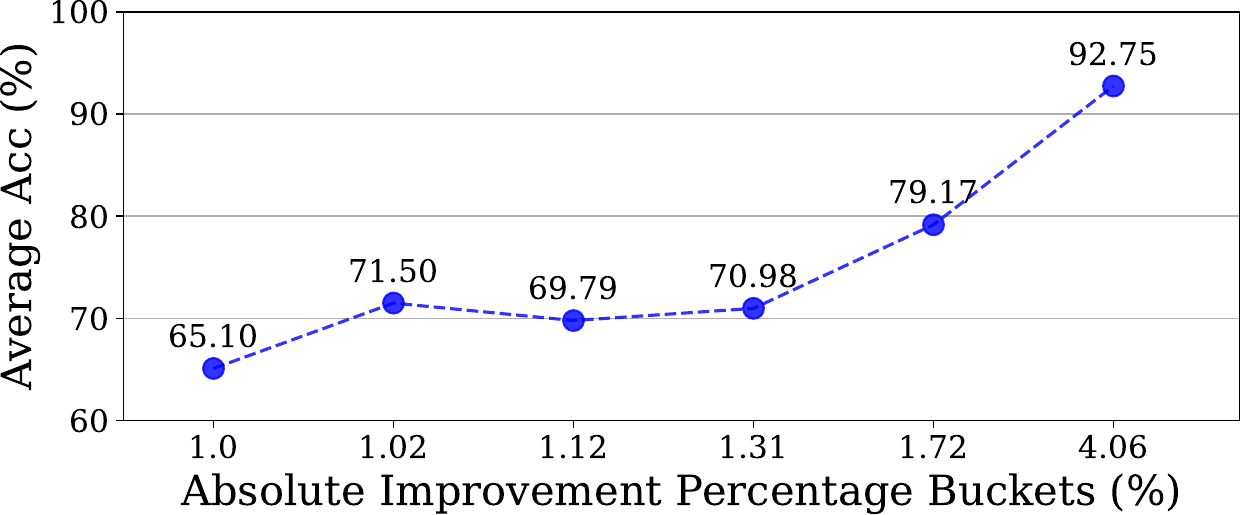}
    \vspace{-0.075in}
    \caption{Results with bucketing the code pairs according to their absolute relative improvement in efficiency, on Python.}
    \label{fig:bucket}
    \vspace{-0.125in}
\end{figure}

\paragraph{Dataset}
To validate the efficacy of our approach to judge code efficiency, we should collect pairs of two different versions of codes before and after modifications. Here, we consider three different scenarios of code editing, and, for the cases where humans refine the code, we use a dataset of code edits made by humans from~\citet{pie-perf}. For the other scenarios where the machine improves the human- or machine-generated codes, we prompt the Code LLM (namely DeepSeek-Coder-Instruct-7B)~\cite{deepseekcoder} to refine the given codes for better efficiency. Specifically, starting with the codes generated by humans from the existing dataset, we generate the machine-refined codes with the Code LLM. In addition, from those machine-generated codes, we similarly prompt the Code LLM to improve them. Through these steps, we can obtain pairs of human-human, human-machine, and machine-machine code versions.

\paragraph{Baselines and Our Model}
In this work, as we tackle a novel problem of judging code efficiency, there are no direct baselines available to compare. Therefore, we turn to compare our approach against the basic models powered by LLMs. Specifically, given a code pair, we perform zero-shot and few-shot prompting with LLMs, to decide which one is more efficient. In addition, we also enhance those strategies with Chain-of-Thought prompting~\cite{CoT}, to elicit the reasoning ability of LLMs with the instruction: "Let's think step by step". For our model, we use the classifier (predicting the class of the efficient code) for main experiments, and provide the performance of the regression model (predicting the relative improvement) during analysis. We use DeepSeek-Coder-Instruct-1.3B for all models, and also provide the results with GPT-3.5, GPT-4o for benchmarking the performance of stronger LLMs without finetuning.

\begin{table}
\caption{Breakdown results for varying the code refinement scenarios. 'H' indicates Human and 'M' indicates Machine.}
\vspace{-0.1in}
\label{tab:scenario_comparison}
\small
\centering
\resizebox{0.48\textwidth}{!}{

\begin{tabular}{lcccccc}
\toprule
  & \multicolumn{3}{c} {\bf Statistics} & \multicolumn{3}{c}{\bf Breakdown Acc }  \\
\cmidrule(l{2pt}r{2pt}){2-4} \cmidrule(l{2pt}r{2pt}){5-7}
Scenarios & Avg. Improve & Degrade \% & Improve \% & Python & C++ & All \\ 

 \midrule
 \midrule
H-H & 1.08 & 37.19 & 21.85 & 80.43 & 64.43 & 67.88 \\
M-M & 1.32 & 30.69 & 32.25 & 60.77 & 55.84 & 57.03 \\ 
H-M & 1.29  & 31.26 & 30.76 & 67.27 & 61.55 & 62.87 \\

\bottomrule

\end{tabular}
}

\end{table}
 
\begin{table}[t!]
\caption{Results with predicting the relative difference of the modified code over its original code in efficiency. Corr denotes the Spearman's rank correlation coefficient, and Acc denotes the accuracy where we convert prediction values into classes.}
\vspace{-0.1in}
\label{tab:main_regression}
\small
\centering
\resizebox{0.4\textwidth}{!}{
\renewcommand{\arraystretch}{0.9}
\begin{tabular}{lcccccc}
\toprule

 & \multicolumn{2}{c}{\bf All} & \multicolumn{2}{c}{\bf Easy }  \\
\cmidrule(l{2pt}r{2pt}){2-3} \cmidrule(l{2pt}r{2pt}){4-5}
 & Corr & Acc  & Corr & Acc \\

\midrule
\midrule

Python & 0.66 & 76.38 & 0.64 & 82.88  \\
C++ & 0.50 & 66.69 & 0.63 & 80.16  \\
Python \& C++ & 0.56 & 68.87 & 0.66 & 81.17 \\

\bottomrule

\end{tabular}
}
\vspace{-0.1in}
\end{table}

\begin{table}[t]
\caption{Generalization results by varying the training data.}
\vspace{-0.05in}
\label{tab:train_result}
\small
\centering
\resizebox{0.375\textwidth}{!}{
\renewcommand{\arraystretch}{1.075}
\begin{tabular}{lccc}
\toprule
 Training Datasets & Python & C++ & All \\
 \midrule
 \midrule
 Python & {\bf 72.75} & 58.26 & 61.52  \\
 C++ & 57.53 & 60.32 & 57.90 \\
 
 \noalign{\vskip 0.25ex}\cdashline{1-4}\noalign{\vskip 0.75ex}

 Python \& C++ & 72.49 & {\bf 62.08} & {\bf 64.42} \\

\bottomrule

\end{tabular}
}

\end{table}

\begin{table}[t]
\caption{The performances of baselines and our model (trained on Python and C++) to other low-resource programming languages (such as Ruby, Perl, and Rust).}
\vspace{-0.05in}
\label{tab:other_pl_results}
\small
\centering
\resizebox{0.475\textwidth}{!}{
\renewcommand{\arraystretch}{1.075}
\begin{tabular}{lcccccccc}
\toprule

 & \multicolumn{4}{c}{\bf All} & \multicolumn{4}{c}{\bf Easy }  \\
\cmidrule(l{2pt}r{2pt}){2-5} \cmidrule(l{2pt}r{2pt}){6-9}
 & Ruby & Perl & Rust& All & Ruby & Perl & Rust& All \\

\midrule
\midrule

Zero-shot & 52.30 & 40.00 & 49.13 & 51.08 & 50.00 & 45.83 & 46.67 & 49.22  \\
Few-shot & 52.06 & 40.00 & 49.57 & 50.99 & 49.01 & 45.83 & 46.67 & 48.44 \\

\noalign{\vskip 0.25ex}\cdashline{1-9}\noalign{\vskip 0.75ex}

Zero-shot CoT & 52.30 & 42.00 & 48.70 & 51.08 & 53.96 & 45.83 & 33.33 & 50.78 \\
Few-shot CoT & 52.06 & 40.00 & 49.57 & 50.99 & 49.01 & 45.83 & 46.67 & 48.44 \\

\noalign{\vskip 0.25ex}\cdashline{1-9}\noalign{\vskip 0.75ex}

\textbf{Ours} & \textbf{58.47} & \textbf{70.00} & \textbf{55.22} & \textbf{58.32} & \textbf{71.78} & \textbf{66.67} & \textbf{60.00} & \textbf{69.92}
 \\

\bottomrule

\end{tabular}
}
\vspace{-0.1in}
\end{table}

\subsection{Experimental Results}

\paragraph{Main Results}
We report the main results in Table~\ref{tab:main_classification}, and, from this, we observe that our method consistently outperforms all baseline models across all settings. Surprisingly, we find that our approach is substantially superior to GPT-3.5 and GPT-4o, demonstrating the continued limitations of even larger models in judging the code efficiency, which further supports the efficacy of our training strategy for it. In addition, our model is particularly effective in scenarios where there is a clear difference in code efficiency — specifically, a difference exceeding 10\% (the easy setting). To examine the performance of our model more granularly based on varying degrees of efficiency differences between code pairs, we bucketize the code pairs based on their efficiency differences. As shown in Figure~\ref{fig:bucket}, the performance of our model increases when the difference between two codes becomes larger. 

\paragraph{Results with Varying Refinement Scenarios}
It is worth noting that our code refinement scenario is categorized as human-human, human-machine, and machine-machine, and we report their breakdown results in Table~\ref{tab:scenario_comparison}. From this, we first observe that the percentage of efficiency improvement in code refinement scenarios is low, which is around 20\% to 30\%, and it is similar to the percentage of efficiency degradation. In addition, the average improvement made by machines is 30\%, meanwhile, the improvement by humans is around 10\%. On the other hand, as shown in the Breakdown Acc column, our model can more effectively identify the code improved by humans (rather than machines).

\paragraph{Relative Improvement Prediction Results}
In addition to classifying the efficient code given the code pair, we can further predict how much it is improved. For this task, we measure the performance of our model, by ranking all the code pairs based on their relative improvements and comparing them with predicted improvements. Also, if the prediction score exceeds 1.00, we classify this case that there is an improvement during refinement. As shown in Table~\ref{tab:main_regression}, we observe both the high rank correlation coefficient and high accuracy, demonstrating the effectiveness of our approach even in this actual improvement prediction setting.

\paragraph{Generalization on Different Languages}
\label{sec:generalization_main}
To see the generalization ability of our approach to different programming languages, we train the model with Python, C++ or both, and measure the performance on Python and C++ as well as relatively low-resource programming languages (such as Ruby, Perl, and Rust). As shown in Table~\ref{tab:train_result} and Table~\ref{tab:other_pl_results}, we observe that the model trained on one language can be generalizable to the other language, perhaps due to the algorithmic similarities in their codes. Also, this results confirm the broader applicability of our approach to various programming languages.

\paragraph{Results with Different Code LLMs}
To see the performance of different Code LLMs in refining the given codes (in terms of efficiency), and to see the performance of our efficiency judgement model trained with the code pairs constructed by different LLMs, we change the code refinement model from DeepSeekCoder (DSC) to recent CodeQwen~\cite{Bai2023QwenTR}, Granite~\cite{Mishra2024GraniteCM}, and GPT-3.5. As shown in Table~\ref{tab:llm_comparison}, we find that DSC and CodeQwen are superior in improving the efficiency of codes when refining them. Yet, the percentage of improvement made by each model is comparable to the percentage of degradation, which supports again our motivation that we should rethink code refinement. Lastly, our model is able to identify the more efficient code among two different versions of models made across different LLMs.

\begin{table}
\caption{With different Code LLMs, we report their average relative improvement (Avg), as well as the percentage of their degradation and improvement in efficiency. Acc (H) and (M) denote the accuracy on human- and machine-generated codes.}
\vspace{-0.1in}
\label{tab:llm_comparison}
\small
\centering
\resizebox{0.48\textwidth}{!}{
\begin{tabular}{lccccc}
\toprule

LLM & Avg & Degrade \% & Improve \%  & Acc (H) & Acc (M)  \\

 \midrule
 \midrule

 DSC & 1.30 & 31.22 & 31.51 & 80.43 & 65.49 \\

 CodeQwen & 1.33 & 28.01 & 29.12 & 80.60 & 58.86 \\
 
 Granite & 1.13 & 21.46 & 24.55 & 76.69 & 48.97 \\
 
 GPT-3.5 & 1.00 & 25.34 & 21.45 & 78.29 & 53.72 \\
 
 \bottomrule

\end{tabular}
}
\vspace{-0.1in}
\end{table}

\paragraph{Results with Larger Models}
We conduct an auxiliary analysis to see how the performance of different methods changes if a model larger than DeepSeek-Coder-Instruct-1.3B (that we use for main experiments) is used. Specifically, we use its 7B model as the base code LLM and then classify the efficient code given code pairs. As shown in Table~\ref{tab:7b_classification}, we observe results similar to those obtained from the smaller 1.3B model, where our model is consistently superior to other baselines.

\begin{figure}[t!]
    \centering
    \includegraphics[width=0.7\linewidth]{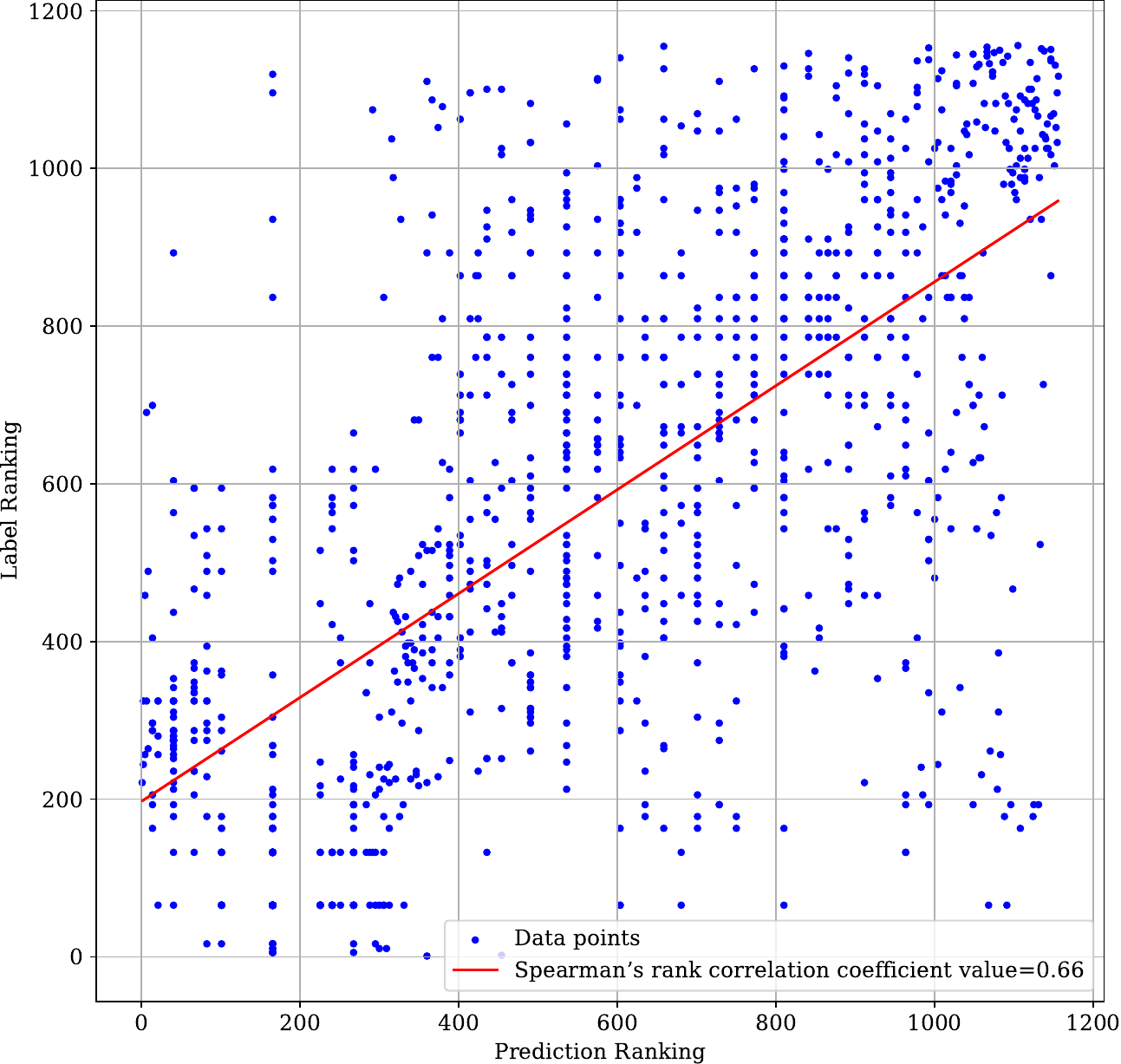}
    \vspace{-0.1in}
    \caption{Visualization of the Spearman’s rank correlation between the ranks of the actual relative improvements and the predicted relative improvements of code pairs, for our model.}
    \label{fig:corr}
    \vspace{-0.15in}
\end{figure}

\paragraph{Visualization of Rank Correlation} 
To visualize how accurate the predicted results of relative improvements of code pairs from our model are, we compare their ranks with the ground-truth ranks calculated by actual relative improvements of code pairs. As shown in Figure~\ref{fig:corr} where we present a scatter plot of rank correlations along with their coefficient value, we observe that the results from our approach have a positive correlation with the ground truth, demonstrating its effectiveness in predicting the relative improvement of code pairs.
\section{Conclusion}
\vspace{-0.05in}
In this work, we pointed out that the codes refined by humans or machines are sometimes inferior than originals, and to tackle this, we introduced a novel approach to identify the more efficient code given a pair of codes before and after modifications. We validated our method on multiple code editing scenarios involving both humans and machines, showcasing its substantial efficacy despite its simplicity.

\begin{table}[t!]
\caption{Results on the DeepSeek-Coder-Instruct-7B model.}
\vspace{-0.1in}
\label{tab:7b_classification}
\small
\centering
\resizebox{0.475\textwidth}{!}{
\renewcommand{\arraystretch}{0.95}
\begin{tabular}{lcccccc}
\toprule

 & \multicolumn{3}{c}{\bf All} & \multicolumn{3}{c}{\bf Easy }  \\
\cmidrule(l{2pt}r{2pt}){2-4} \cmidrule(l{2pt}r{2pt}){5-7}
 & Python & C++ & All & Python & C++ & All \\

\midrule
\midrule

Zero-shot & 52.50 & 47.91 & 48.88 & 51.62 & 49.05 & 49.98   \\
Few-shot & 54.50 & 49.86 &  50.90 & 55.87 & 49.81 & 52.00  \\

\noalign{\vskip 0.25ex}\cdashline{1-7}\noalign{\vskip 0.75ex}

Zero-shot CoT & 54.50 & 49.86 & 50.90 & 55.53 & 52.73 & 53.75  \\
Few-shot CoT & 53.11 & 49.99 &  50.69 & 47.04 & 52.16 & 50.3   \\

\noalign{\vskip 0.25ex}\cdashline{1-7}\noalign{\vskip 0.75ex}

\textbf{Ours} & \textbf{73.44} & \textbf{62.90} & \textbf{65.27} & \textbf{78.88} & \textbf{72.68} & \textbf{74.93}
 \\

\bottomrule

\end{tabular}
}
\end{table}

\section*{Limitation}
\vspace{-0.05in}

There are some areas that future work may improve upon. First, we perform experiments with two widely used programming languages, such as Python and C++, and it may be promising to consider other languages, particularly those used less frequently. In addition, in terms of measuring the code efficiency, we consider its execution time, meanwhile, there may be additional factors to consider, such as memory usage, I/O operations, and the underlying OS environment. Future work may incorporate these factors to perform a more holistic assessment of program efficiency. Lastly, beyond predicting the efficient code, future work may explore its interpretability, providing the reasons why certain codes are more efficient than others at a more fine-grained level (e.g., lines of codes).

\section*{Ethic Statement}
\vspace{-0.05in}

We believe this work does not have particular concerns about ethics. This is because, it strictly focuses on the technological aspect of comparing code efficiency, which does not engage in the unethical use of LLMs for manipulating software codes, user data, or any other sensitive information.

\bibliography{custom}

\appendix
\clearpage

\label{sec:appendix}

\section{Additional Experimental Setups}
\label{appendix:setups}

\begin{table}[t]
\caption{Dataset statistics. The first two rows represent the code statistics made by humans and the other rows are the ones made by machines. PL denotes the programming language.}
\vspace{-0.075in}
\label{tab:train_val_test}
\small
\centering
\resizebox{0.475\textwidth}{!}{
\renewcommand{\arraystretch}{1.075}
\begin{tabular}{lcccc}
\toprule
  & PL & Training & Validation & Test \\
 \midrule
 \midrule
 Human  & Python & 14936 & 761 & 562  \\
 Human & C++ & 27005 & 1316 & 2038  \\
 \noalign{\vskip 0.25ex}\cdashline{1-5}\noalign{\vskip 0.75ex}
 
 DSC & Python & 17079 & 1046 & 594  \\
 DSC & C++ & 35666 & 1290 & 1949 \\

 gpt-3.5 & Python & 10674 & 1306 & 724  \\
  CodeQwen & Python & 14854 & 227 & 207 \\
  Granite & Python & 20783 & 650 & 315 \\

\bottomrule

\end{tabular}
}

\end{table}

\paragraph{Dataset Details}
We report the dataset statistics in Table~\ref{tab:train_val_test}. Note that, in order to obtain the stable code execution result to decide which code is more efficient than the other, we run every refined code with its original code three times and then select the one whose results are consistent across those three runs. In addition, the code execution is performed following the existing setup~\cite{pie-perf}.

\paragraph{Fine-tuning Details}
We provide details on fine-tuning the efficiency judgment model: we fine-tune the Code LLM (namely, DeepSeek-Coder-Instruct-1.3B) over 10 epochs with a batch size of 16 and a learning rate of 2e-5, and we select the best epoch based on performance on the validation set.

\paragraph{Prompts}
In Table~\ref{tab:prompt}, we provide the prompts used to elicit the Code LLM to refine the code and to predict the code efficiency (in classification and regression settings). For the efficiency prediction problem, we randomly shuffle the sequence of the original and its refined codes.

\section{Additional Experimental Results}
\label{appendix:results}
Here, we provide additional experimental results.

\begin{table}[t!]
\caption{Average accuracy results for code improvement classification with order perturbation across multiple different runs, where we report the variance in parentheses.}
\vspace{-0.1in}
\label{tab:main_classification_stat}
\small
\centering
\resizebox{0.475\textwidth}{!}{
\begin{tabular}{lcccccc}
\toprule

 & \multicolumn{3}{c}{\bf All} & \multicolumn{3}{c}{\bf Easy }  \\
\cmidrule (l{2pt}r{2pt}){2-4} \cmidrule (l{2pt}r{2pt}){5-7}
 & Python & C++ & All & Python & C++ & All \\

\midrule
\midrule

Zero-shot & 50.26 (0.7) & 47.02 (5.9) & 47.75 (2.9) & 48.49 (4.2) & 49.94 (3.2) & 49.42 (0.2) \\ 
Few-shot & 50.43 (1.2) & 50.00 (2.7) & 50.10 (2.3) & 49.61 (0.4) & 48.32 (0.0) & 48.79 (0.1) \\

\noalign{\vskip 0.25ex}\cdashline{1-7}\noalign{\vskip 0.75ex}

Zero-shot CoT & 49.70 (0.9) & 50.53 (2.7) & 50.34 (2.2) & 48.94 (0.0) & 50.29 (6.6) & 49.80 (2.7) \\
Few-shot CoT & 49.27 (3.2) & 49.79 (2.3) & 49.67 (2.5) & 50.62 (2.5) & 48.03 (0.0) & 48.97 (0.3) \\

\noalign{\vskip 0.25ex}\cdashline{1-7}\noalign{\vskip 0.75ex}

\textbf{Ours} & \textbf{71.67} (1.3) & \textbf{62.09} (0.0) & \textbf{64.25} (0.1) & \textbf{77.60} (0.0) & \textbf{70.52} (0.3) & \textbf{73.09} (0.1) \\

\bottomrule

\end{tabular}
}
\end{table}

\paragraph{Analysis on Bias for Code Sequence}
In our code efficiency judgment task, we put a sequence of two codes in the input of Code LLMs, and the Code LLMs may have a bias in this sequence (e.g., predicting the code at the last more often). To see whether they have such a bias, we conduct an additional experiment, flipping the order of the code pairs in the input. We report the results in Table~\ref{tab:main_classification_stat}, and, from this, we observe that there are no such the notable bias in the sequence of codes.

\begin{table}[t] 
\caption{The performance difference between the cases of two classes (Degrade and Improve) and three classes (Degrade, Non-Improve, and Improve), with the regression model.}
\vspace{-0.075in}
\label{tab:two_classes_result}
\small
\centering
\resizebox{0.375\textwidth}{!}{
\renewcommand{\arraystretch}{1.075}
\begin{tabular}{lcc}
\toprule
 Training Datasets & Two Classes & Three Classes \\
 \midrule
 \midrule
 Python & 76.38 & 65.40  \\
 C++ & 66.69 & 70.00 \\
 
 \noalign{\vskip 0.25ex}\cdashline{1-3}\noalign{\vskip 0.75ex}

 Python \& C++ & 68.87 & 68.97 \\

\bottomrule

\end{tabular}
}

\vspace{-0.05in}
\end{table}

\paragraph{Qualitative Analysis}
We provide some example codes in Python and C++ in Figures~\ref{fig:example} and~\ref{fig:example2}. From these two examples, we observe that, despite the difference in grammar across different programming languages, code pairs from them can share the same underlying algorithms. This result supports our finding on generalization ability that our model trained on one programming language can be generalizable to other languages (See Table~\ref{tab:train_result}).

\paragraph{Two Classes Analysis}
For an auxiliary analysis, we measure the performance of the proposed regression model on the dataset with three different prediction classes of Degrade, Non-Improve, and Improve, unlike the one with two classes of Degrade and Improve in our main experiment. As shown in~\ref{tab:two_classes_result}, we observe that the overall performance between two classes (Degrade and Improve) and three classes (Degrade, Non-Improve, and Improve) cases is similar. Furthermore, for Python, the performance decreases with the case of three classes, meanwhile, for C++, the performance increases that may be due to the fact that most code pairs for C++ belong to the Non-Improve class (Figure~\ref{fig:main_figure}) and our model accurately identifies this.

\begin{table*}[t!]
    \caption{A list of prompts that we used for code refinement and efficiency predictions. It is worth noting that the variable inside the parentheses \{\} is replaced with its actual code.}
    \label{tab:prompt}
    \vspace{-0.075in}
    \resizebox{1\textwidth}{!}{
        \begin{tabular}{ll}
        \toprule
        \multicolumn{1}{p{.3\textwidth}}{\textbf{Types}} & \makecell{\multicolumn{1}{p{.7\textwidth}}{\textbf{Prompts}}} \\
        \midrule
        \multicolumn{1}{p{.3\textwidth}}{\textbf{Code Refinement}} & 
        \makecell{
            \multicolumn{1}{p{.7\textwidth}}{Update the given code to make it more efficient. \{Original code\}} 
        }\\
        \midrule
        \multicolumn{1}{p{.3\textwidth}}{\textbf{Efficiency Classification}} & 
        \makecell{
            \multicolumn{1}{p{.7\textwidth}}{Given a selection of code, determine which one is the most efficient in computing.} \\
            \multicolumn{1}{p{.7\textwidth}}{A: \{Original code or Refined code\}} \\
            \multicolumn{1}{p{.7\textwidth}}{B: \{Refined code or Original code\}}
        } \\
                
\noalign{\vskip 0.25ex}\cdashline{1-2}\noalign{\vskip 0.75ex}
         \multicolumn{1}{p{.3\textwidth}}{\textbf{Efficiency Regression}} &
        \makecell{
\multicolumn{1}{p{.7\textwidth}}{Given two sets of code, assess how much Code B has improved compared to Code A.} \\
\multicolumn{1}{p{.7\textwidth}}{A. \{Original code\}} \\
\multicolumn{1}{p{.7\textwidth}}{B. \{Refined code\}} \\
}\\
        \bottomrule
        \end{tabular}
    }
\end{table*}
\begin{figure*}[t!]
    \begin{minipage}{\textwidth}
    \begin{lstlisting}[language=Python, numbers=none]
# Python Example
N = int(eval(input()))
print(((N*(N-1))//2))
\end{lstlisting}

    \begin{lstlisting}[language=C++, numbers=none]
// C++ Example
#include<iostream>
using namespace std;
int main() {
        long long int n;
        cin>>n;
        cout<<n*(n-1)/2<< endl;
}
\end{lstlisting}

    \end{minipage}
    \vspace{-0.175in}
    \caption{Generated Python and C++ samples for the question  "For an integer N, we will choose a permutation $\{P_1, P_2, ..., P_N\}$ of $\{1, 2, ..., N\}$. Then, for each $i=1,2,...,N,$ let $M_i$ be the remainder when i is divided by $P_i$. Find the maximum possible value of $M_1 + M_2 + \cdots + M_N$. Constraints $N$ is an integer satisfying $1 \leq N \leq 10^9$".}

    \label{fig:example}
\end{figure*}

\begin{figure*}[t!]
    \begin{minipage}{\textwidth}
    \begin{lstlisting}[language=Python, numbers=none]
# Python Example
from math import floor,ceil

X = int(eval(input()))
cash = 100
count = 0
while cash < X:
    cash=floor(cash*1.01)
    count += 1

print(count)
\end{lstlisting}

    \begin{lstlisting}[language=C++, numbers=none]
// C++ Example
#include <bits/stdc++.h>
using namespace std;
int main() {
    long long X;
    cin >> X;

    int year=0;
    long long s=100;

    while(s<X){
        s=s*1.01;       
        year++;       
    }
    cout << year << endl;
}
\end{lstlisting}

    \end{minipage}
    \vspace{-0.175in}
    \caption{Generated Python and C++ codes for the question  "Takahashi has a deposit of 100 yen (the currency of Japan) in AtCoder Bank. The bank pays an annual interest rate of 1\% compounded annually. (A fraction of less than one yen is discarded.) Assuming that nothing other than the interest affects Takahashi's balance, in how many years does the balance reach X yen or above for the first time?".}

    \label{fig:example2}
\end{figure*}

\begin{figure*}
\small
\begin{spacing}{1.0}
\begin{mybox}{User}

We compared the efficiency in computing of codes A and B. As a result, code A was more efficient than B. The improvement fraction is 1.7. Explain why this might be the case.

\begin{lstlisting}[language=Python]
# Code A
n = input()
total = 0
for i in range(2, n+1, 2):
    total += i
\end{lstlisting}
\vspace{-0.1in}
\begin{lstlisting}[language=Python]
# Code B
n = input()
total = 0
for i in range(1, n+1):
    if i % 2 == 0:
        total += i
\end{lstlisting}
\end{mybox}

\begin{mybox}{Response}

The efficiency comparison between Code A and Code B, and why Code A is 1.7 times more efficient than Code B, can be explained based on how each code is structured to perform its task. The task here is to compute the sum of all even numbers up to a given nnn. Here's a breakdown of the differences:\\

\vspace{-0.1in}
\begin{lstlisting}[language=Python]
# Code A
n = input()
total = 0
for i in range(2, n+1, 2):
    total += i
\end{lstlisting}
\vspace{-0.1in}
\begin{itemize}[itemsep=1.0mm, parsep=1pt, leftmargin=*]
    \item Range Function: This code utilizes the \lstinline|range()| function with three arguments: the start, stop, and step. It starts at 2 and increments by 2 each iteration, effectively looping only over even numbers.
    \item Loop Count: The number of iterations in this loop is directly halved compared to Code B, because it increments by 2. For a maximum number nnn, this loop executes approximately n/2n/2n/2 times. 
    \item Conditional Checks: There are no conditional checks inside the loop. Each iteration unconditionally adds the loop variable iii (which is guaranteed to be even) to \lstinline|total|.
\end{itemize}
\vspace{-0.1in}
\begin{lstlisting}[language=Python]
# Code B
n = input()
total = 0
for i in range(1, n+1):
    if i % 2 == 0:
        total += i
\end{lstlisting}
\vspace{-0.1in}
\begin{itemize}[itemsep=1.0mm, parsep=1pt, leftmargin=*]
    \item Range Function: Here, the \lstinline|range()| function is used with just the start and stop, defaulting to a step of 1. It iterates over every integer from 1 to nnn.
    \item Loop Count: The loop iterates nnn times, which is twice as many as Code A for the same nnn.
    \item Conditional Checks: Inside the loop, there is a conditional check \lstinline|(if i % 2 == 0)| to determine whether each number is even before adding it to total. This conditional check adds computational overhead for each iteration.
\end{itemize}

\textbf{Why Code A is More Efficient}
\begin{enumerate}[itemsep=1.0mm, parsep=1pt, leftmargin=*]
  \item Fewer Iterations: Code A runs fewer iterations—only about half as many as Code B. This reduces the number of operations (additions and assignments) executed.
  \item No Conditional Logic: Code A does not use any conditional logic inside its loop, which saves the time required for evaluating the condition for each iteration in Code B.
  \item Optimized Looping: Directly iterating over only the numbers of interest (even numbers) avoids unnecessary iterations over odd numbers, which Code B does but doesn't process.
\end{enumerate}
The improvement fraction of 1.7 suggests that Code A is about 70\% more efficient than Code B. This could be measured in terms of execution time or the number of operations performed, and it results largely from the reduced number of iterations and the elimination of conditional checks within each iteration.
\end{mybox}
\vspace{-0.15in}

\caption{A toy example of an explainability of our approach.}
\end{spacing}

\label{fig:toy_example1}
\end{figure*}

\begin{figure*}
\small
\begin{spacing}{1.0}

\begin{mybox}{User}

We compared the efficiency in computing of codes A and B. As a result, code B was more efficient than A. The improvement fraction is 3.02. Explain why this might be the case.

\vspace{0.1in}
\begin{lstlisting}[language=Python]
# Code A
from math import factorial as f
n = int(eval(input()))
print((f(n) % (10 ** 9 + 7)))
\end{lstlisting}
\vspace{-0.1in}
\begin{lstlisting}[language=Python]
# Code B
def factorial(n, mod=10**9 + 7):
    f = [0]*(n+1)\n f[0] = 1
    for i in range(1, n+1):
        f[i] = (f[i-1]*i) % mod
    return f[n]
n = int(input())
print(factorial(n))
\end{lstlisting}
\end{mybox}

\begin{mybox}{Response}

The improvement in efficiency from code A to code B by a factor of 3.02 can be attributed to several reasons, primarily concerning how each code handles the calculation of the factorial and the management of large integers. \\

\textbf{1. Handling of Large Numbers} \\
Code A uses Python's built-in \lstinline|factorial| function from the \lstinline|math| library. This function computes the factorial of n as a single large integer, which can become extremely large as n increases. Only after calculating this massive number does it apply the modulo operation. This approach is memory and computation-intensive because it deals with very large numbers up to the final computation.\\
Code B takes a different approach by iteratively calculating the factorial and immediately applying the modulo operation at each step. This significantly reduces the size of the integers the program must handle at any one time. By ensuring that no intermediate value exceeds \lstinline|10**9 + 7|, Code B avoids the overhead associated with manipulating very large numbers.\\

\textbf{2. Efficiency in Modulo Operation} \\
Immediate Modulo Application: In Code B, by applying the modulo operation in each step (\lstinline|f[i] = (f[i-1]*i) % mod|), the size of the numbers involved in the calculations remains bounded by mod. This is a more efficient use of space and typically faster to compute, as each operation is simpler and avoids the potential overhead of managing and operating on very large integers. \\

\textbf{3. Space Complexity} \\ 
Code B also explicitly manages an array f which stores all the factorial values modulo \lstinline|mod| up to \lstinline|n|. This allows not only for efficient calculation of \lstinline|n! % mod| but potentially could be adapted to handle multiple queries more efficiently by reusing precomputed factorials if needed.\\

\textbf{4. Use of Python's Built-in vs Custom Function} \\
Optimization: Python's built-in functions are highly optimized; however, when it comes to specific tasks like calculating a factorial modulo a number, a custom-tailored approach like in Code B can outperform by optimizing for the particular constraints (like the modulo operation being used after each multiplication). \\

\textbf{5. Potential Overhead in Code A} \\
The \lstinline|eval| function used in Code A is generally slower and less safe than a direct \lstinline|input()| function as used in Code B. While the impact on performance here is minimal compared to the factorial calculation itself, it's another aspect where Code B is more optimized.
In summary, Code B is more efficient due to its continuous management of number size through the modulo operation, reducing the computational and memory overhead by avoiding the manipulation of extremely large numbers. Additionally, the direct approach without the overhead of unnecessary functions (like \lstinline|eval|) contributes to its higher efficiency. \\

\end{mybox}
\vspace{-0.15in}
\label{fig:toy_example2}
\caption{A toy example of an explainability of our approach.}
\end{spacing}
\end{figure*}

\end{document}